\documentstyle [12pt] {article}

\catcode`\@=11
\def\@citex[#1]#2{%
\if@filesw \immediate \write \@auxout {\string \citation {#2}}\fi
\@tempcntb\m@ne \let\@h@ld\relax \def\@citea{}%
\@cite{%
  \@for \@citeb:=#2\do {%
    \@ifundefined {b@\@citeb}%
      {\@h@ld\@citea\@tempcntb\m@ne{\bf ?}%
      \@warning {Citation `\@citeb ' on page \thepage \space
undefined}}%
      {\@tempcnta\@tempcntb \advance\@tempcnta\@ne%
      \@tempcntb\number\csname b@\@citeb \endcsname \relax%
      \ifnum\@tempcnta=\@tempcntb 
	\ifx\@h@ld\relax%
	  \edef \@h@ld{\@citea\csname b@\@citeb\endcsname}%
	\else%
	  \edef\@h@ld{\ifmmode{-}\else--\fi\csname
b@\@citeb\endcsname}%
	\fi%
      \else
	\@h@ld\@citea\csname b@\@citeb \endcsname%
	\let\@h@ld\relax%
      \fi}%
    \def\@citea{,\penalty\@highpenalty\,}%
  }\@h@ld
}{#1}}

\def\@citeb#1#2{{[#1]\if@tempswa , #2\fi}}
\def\@citeu#1#2{{$^{#1}$\if@tempswa , #2\fi }}
\def\@citep#1#2{{#1\if@tempswa , #2\fi}}

\def\bcites{         
	\catcode`\@=11
	\let\@cite=\@citeb
	\catcode`\@=12
}

\def\upcites{         
	\catcode`\@=11
	\let\@cite=\@citeu
	\catcode`\@=12
}

\def\plaincites{      
	\catcode`\@=11
	\let\@cite=\@citep
	\catcode`\@=12
}

\newcount\hour
\newcount\minute
\newtoks\amorpm
\hour=\time\divide\hour by 60
\minute=\time{\multiply\hour by 60 \global\advance\minute by-\hour}
\edef\standardtime{{\ifnum\hour<12 \global\amorpm={am}%
	\else\global\amorpm={pm}\advance\hour by-12 \fi
	\ifnum\hour=0 \hour=12 \fi
	\number\hour:\ifnum\minute<10
0\fi\number\minute\the\amorpm}}
\edef\militarytime{\number\hour:\ifnum\minute<10
0\fi\number\minute}

\def\draftlabel#1{{\@bsphack\if@filesw {\let\thepage\relax
   \xdef\@gtempa{\write\@auxout{\string
      \newlabel{#1}{{\@currentlabel}{\thepage}}}}}\@gtempa
   \if@nobreak \ifvmode\nobreak\fi\fi\fi\@esphack}
	\gdef\@eqnlabel{#1}}
\def\@eqnlabel{}
\def\@vacuum{}
\def\marginnote#1{}
\def\draftmarginnote#1{\marginpar{\raggedright\scriptsize\tt#1}}
\def\draft{
	\pagestyle{plain}
	\overfullrule=2pt
	\oddsidemargin -.5truein
	\def\@oddhead{\sl \phantom{\today\quad\militarytime} \hfil
	\smash{\Large\sl DRAFT} \hfil \today\quad\militarytime}
	\let\@evenhead\@oddhead
	\let\label=\draftlabel
	\let\marginnote=\draftmarginnote
	\def\ps@empty{\let\@mkboth\@gobbletwo
	\def\@oddfoot{\hfil \smash{\Large\sl DRAFT} \hfil}
	\let\@evenfoot\@oddhead}

\def\@eqnnum{(\theequation)\rlap{\kern\marginparsep\tt\@eqnlabel}%
	\global\let\@eqnlabel\@vacuum}  }

\def\blackfonts{
	\font\blackboard=msbm10 scaled\magstep1
	\font\blackboards=msbm8
	\font\blackboardss=msbm6
}

\def\nblack{            
	\def\ZZ{{Z \n{10} Z}}
	\def\NN{{N \n{14} N}}
	\def\CC{{C \n{11} C}}
	\def\RR{{R \n{11} R}}
	\def\QQ{{Q \n{12} Q}}
	\def\PP{{P \n{11} P}}
}

\def\prep{         
	\catcode`\@=11
	\input art10.sty
	\catcode`\@=12
	
	\let\small\null
	\def\blackfonts{
		\font\blackboard=msbm10
		\font\blackboards=msbm7
		\font\blackboardss=msbm5
	}
	\let\sl\it
	\twocolumn
	\sloppy
	\voffset=-2.54truecm
	\hoffset=-2.54truecm
	\flushbottom
	\parindent 1em
	\leftmargini 2em
	\leftmarginv .5em
	\leftmarginvi .5em
	\marginparwidth 48pt
	\marginparsep 10pt
	\setlength{\columnsep}{2truecm}
	\setlength{\textwidth}{25.4truecm}
	\setlength{\textheight}{17truecm}
	\baselineskip=16pt
	\oddsidemargin .18truein
	\evensidemargin .17truein
}

\def\eqalign#1{\null\,\vcenter{\openup\jot\m@th
  \ialign{\strut\hfil$\displaystyle{##}$&$\displaystyle{{}##}$\hfil
      \crcr#1\crcr}}\,}
\def\eqalignno#1{\displ@y \tabskip\centering
  \halign
to\displaywidth{\hfil$\@lign\displaystyle{##}$\tabskip\z@skip
    &$\@lign\displaystyle{{}##}$\hfil\tabskip\centering
    &\llap{$\@lign##$}\tabskip\z@skip\crcr
    #1\crcr}}

\def\section{\@startsection {section}{1}{\z@}{3.ex plus 1ex minus
 .2ex}{2.ex plus .2ex}{\large\bf}}
\def\subsection{\@startsection{subsection}{2}{\z@}{2.75ex plus 1ex
minus
 .2ex}{1.5ex plus .2ex}{\bf}}

\def\appendix{{\newpage\section*{Appendix}}\let\appendix\section%
	{\setcounter{section}{0}
	\gdef\thesection{\Alph{section}}}\section}

\def\abstract{\if@twocolumn
\section*{Abstract}
\else 
\begin{center}
{\bf Abstract\vspace{-.5em}\vspace{0pt}}
\end{center}
\quotation
\fi}

\catcode`\@=12

\def\ibid{{\it ibid.\/}}

\newcommand{\beq}{\begin{equation}}
\newcommand{\eeq}{\end{equation}}
\newcommand{\beqa}{\begin{eqnarray}}
\newcommand{\eeqa}{\end{eqnarray}}

\def\noj#1,#2,{{\bf #1} (19#2)\ }
\def\jou#1,#2,#3,{{\sl #1\/ }{\bf #2} (19#3)\ }
\def\ann#1,#2,{{\sl Ann.\ Physics\/ }{\bf #1} (19#2)\ }
\def\cmp#1,#2,{{\sl Comm.\ Math.\ Phys.\/ }{\bf #1} (19#2)\ }
\def\ma#1,#2,{{\sl Math.\ Ann.\/ }{\bf #1} (19#2)\ }
\def\jd#1,#2,{{\sl J.\ Diff.\ Geom.\/ }{\bf #1} (19#2)\ }
\def\invm#1,#2,{{\sl Invent.\ Math.\/ }{\bf #1} (19#2)\ }
\def\cq#1,#2,{{\sl Class.\ Quantum Grav.\/ }{\bf #1} (19#2)\ }
\def\cqg#1,#2,{{\sl Class.\ Quantum Grav.\/ }{\bf #1} (19#2)\ }
\def\ijmp#1,#2,{{\sl Int.\ J.\ Mod.\ Phys.\/ }{\bf A#1} (19#2)\ }
\def\jmphy#1,#2,{{\sl J.\ Geom.\ Phys.\/ }{\bf #1} (19#2)\ }
\def\jams#1,#2,{{\sl J.\ Amer.\ Math.\ Soc.\/ }{\bf #1} (19#2)\ }
\def\grg#1,#2,{{\sl Gen.\ Rel.\ Grav.\/ }{\bf #1} (19#2)\ }
\def\mpl#1,#2,{{\sl Mod.\ Phys.\ Lett.\/ }{\bf A#1} (19#2)\ }
\def\nc#1,#2,{{\sl Nuovo Cim.\/ }{\bf #1} (19#2)\ }
\def\np#1,#2,{{\sl Nucl.\ Phys.\/ }{\bf B#1} (19#2)\ }
\def\pl#1,#2,{{\sl Phys.\ Lett.\/ }{\bf #1B} (19#2)\ }
\def\pla#1,#2,{{\sl Phys.\ Lett.\/ }{\bf #1A} (19#2)\ }
\def\pr#1,#2,{{\sl Phys.\ Rev.\/ }{\bf #1} (19#2)\ }
\def\prd#1,#2,{{\sl Phys.\ Rev.\/ }{\bf D#1} (19#2)\ }
\def\prl#1,#2,{{\sl Phys.\ Rev.\ Lett.\/ }{\bf #1} (19#2)\ }
\def\prp#1,#2,{{\sl Phys.\ Rept.\/ }{\bf #1C} (19#2)\ }
\def\ptp#1,#2,{{\sl Prog.\ Theor.\ Phys.\/ }{\bf #1} (19#2)\ }
\def\ptpsup#1,#2,{{\sl Prog.\ Theor.\ Phys.\/ Suppl.\/ }{\bf #1}
(19#2)\ }
\def\rmp#1,#2,{{\sl Rev.\ Mod.\ Phys.\/ }{\bf #1} (19#2)\ }
\def\yadfiz#1,#2,#3[#4,#5]{{\sl Yad.\ Fiz.\/ }{\bf #1} (19#2) #3%
\ [{\sl Sov.\ J.\ Nucl.\ Phys.\/ }{\bf #4} (19#2) #5]}
\def\zh#1,#2,#3[#4,#5]{{\sl Zh..\ Exp.\ Theor.\ Fiz.\/ }{\bf #1}
(19#2) #3%
\ [{\sl Sov.\ Phys.\ JETP\/ }{\bf #4} (19#2) #5]}

\def\beq{\begin{equation}}
\def\eeq{\end{equation}}
\def\beqar{\begin{eqnarray}}
\def\eeqar{\end{eqnarray}}

\def\nfrac#1#2{{\displaystyle{\vphantom1\smash{\lower.5ex\hbox{\sma
ll$#1$}}%
	\over\vphantom1\smash{\raise.25ex\hbox{\small$#2$}}}}}

\def\p#1{\mskip#1mu}
\def\n#1{\mskip-#1mu}
\def\stop{\p6.}
\def\comma{\p6,}

\def\lae{\mathrel{\mathop{\smash{\lower .5 ex
\hbox{$\stackrel<\sim$}}}}}
\def\lae{\mathrel{\mathop{\smash{\lower .5 ex
\hbox{$\stackrel>\sim$}}}}}

\def\l:{\mathopen{:}\,}
\def\r:{\,\mathclose{:}}

\catcode`\@=11
\def\theequation{\arabic{equation}}
\catcode`\@=12

\nblack
\bcites

\nblack

\catcode`\@=11
\def\theequation{\thesection.\arabic{equation}}
\@addtoreset{equation}{section}
\@addtoreset{footnote}{section}
\@addtoreset{footnote}{subsection}
\catcode`\@=12

\newcommand{\beqn}{\begin{equation}}
\newcommand{\eeqn}{\end{equation}}
\newcommand{\beqnarray}{\begin{eqnarray}}
\newcommand{\eeqnarray}{\end{eqnarray}}
\newcommand {\bear} [1] {\begin {array} {#1}}
\newcommand {\ear} {\end {array}}

\newcommand {\beqarn} {\begin{eqnarray*}}
\newcommand {\eeqarn} {\end{eqnarray*}}

\newcommand {\mrm} [1] {\mathrm {#1}}

\newcommand {\tr} {{\mbox {$\mrm{tr}$}}}

\newcommand {\myref} [1]	%
	{%
	\begin{thebibliography} {99}	%
			{#1}	%
	\end {thebibliography}}

\def\emph#1{{\em #1}}
\def\mathbf#1{{\bf #1}}
\def\mathrm#1{{\rm #1}}
\def\mathit#1{{\it #1}}

\catcode`\@=11
\@addtoreset{footnote}{section}
\catcode`\@=12

\newcounter	{exinsert}	[subsection]

\renewcommand {\theexinsert}	%
{	\thesubsection.\arabic{equation}}

\renewcommand {\theequation} {\thesection.\arabic{equation}}

\catcode`\@=11
\@addtoreset{equation}{section}
\catcode`\@=12

\newenvironment {exinsert} [1]	%
{	
	\begin {quotation}	%
	\refstepcounter {equation}	%
	{\bf {#1}} \theequation. \,\,	%
}
{	
	\end {quotation}
}

\newcommand {\bprop}
{\begin {exinsert} {Proposition}
}
\newcommand {\eprop} {\end {exinsert} }

\newcommand {\bexe} {\begin {exinsert} {Exercise}}
\newcommand {\eexe} {\end {exinsert} }

\newcommand {\bexa} {\begin {exinsert} {Example}}
\newcommand {\eexa} {\end {exinsert} }

\newcommand {\literal} [1] {{ \mbox {$\mrm {#1}$}}}

\newcommand {\grad} {\nabla}

\newcommand {\ncmd} [2] {\newcommand {#1} {#2}}

\ncmd {\vgrad} {{\vec \grad}}

\ncmd {\Mv} {{\cal M}_V}
\ncmd {\Mh} {{\cal M}_H}

\ncmd {\adj} {\literal {adj}}	
\ncmd {\Mp} {M_\literal {planck}}	

\ncmd {\x} {$\times$}

\newcommand {\btab} [3]
  {\begin{table} [htb]
   \begin{center}
   \caption {#2}	\label {tab:#1}
   \begin {tabular} {#3}
   \hline}

\newcommand {\etab}
	{  \hline   \end {tabular}
   \end{center}
  \end{table}}

\ncmd {\bR} {\textbf {R}}
\ncmd {\bC} {\textbf {C}}

\ncmd {\CQ} {\cal Q}

\def\jou#1,#2,#3,{{\sl #1\/ }{\bf #2} (19#3)\ }
\def\ibid#1,#2,{{\sl ibid.\/ }{\bf #1} (19#2)\ }
\def\am#1,#2,{{\sl Acta. Math.\/ } {\bf #1} (19#2)\ }
\def\annp#1,#2,{{\sl Ann.\ Physics\/ }{\bf #1} (19#2)\ }
\def\cmp#1,#2,{{\sl Comm.\ Math.\ Phys.\/ }{\bf #1} (19#2)\ }
\def\cqg#1,#2,{{\sl Class.\ Quantum Grav.\/ }{\bf #1} (19#2)\ }
\def\dm#1,#2,{{\sl Duke\ Math.\ J.\/ }{\bf #1} (19#2)\ }
\def\ijmpa#1,#2,{{\sl Int.\ J.\ Mod.\ Phys.\/ }{\bf A#1} (19#2)\ }
\def\invm#1,#2,{{\sl Invent.\ Math.\/ }{\bf #1} (19#2)\ }
\def\jhep#1,#2,{{\sl JHEP\/ }{\bf #1} (19#2)\ }
\def\jmp#1,#2,{{\sl J.\ Geom.\ Phys.\/ }{\bf #1} (19#2)\ }
\def\jams#1,#2,{{\sl J.\ Diff.\ Geom.\/ }{\bf #1} (19#2)\ }
\def\jdg#1,#2,{{\sl J.\ Amer.\ Math.\ Soc.\/ }{\bf #1} (19#2)\ }
\def\jpa#1,#2,{{\sl J.\ Phys.\ A.\/ }{\bf #1} (19#2)\ }
\def\grg#1,#2,{{\sl Gen.\ Rel.\ Grav.\/ }{\bf #1} (19#2)\ }
\def\mpla#1,#2,{{\sl Mod.\ Phys.\ Lett.\/ }{\bf A#1} (19#2)\ }
\def\ma#1,#2,{{\sl Math.\ Ann.\/ } {\bf #1} (19#2)\ }
\def\nc#1,#2,{{\sl Nuovo\ Cim.\/ }{\bf #1} (19#2)\ }
\def\npb#1,#2,{{\sl Nucl.\ Phys.\/ }{\bf B#1} (19#2)\ }
\def\plb#1,#2,{{\sl Phys.\ Lett.\/ }{\bf #1B} (19#2)\ }
\def\pla#1,#2,{{\sl Phys.\ Lett.\/ }{\bf #1A} (19#2)\ }
\def\pnas#1,#2,{{\sl Proc.Nat.Acad.Sci.\/ }{\bf #1} (19#2)\ }
\def\prev#1,#2,{{\sl Phys.\ Rev.\/ }{\bf #1} (19#2)\ }
\def\prd#1,#2,{{\sl Phys.\ Rev.\/ }{\bf D#1} (19#2)\ }
\def\prl#1,#2,{{\sl Phys.\ Rev.\ Lett.\/ }{\bf #1} (19#2)\ }
\def\prpt#1,#2,{{\sl Phys.\ Rept.\/ }{\bf #1C} (19#2)\ }
\def\ptp#1,#2,{{\sl Prog.\ Theor.\ Phys.\/ }{\bf #1} (19#2)\ }
\def\ptpsup#1,#2,{{\sl Prog.\ Theor.\ Phys.\/ Suppl.\/ }{\bf #1} (19#2)\ }
\def\rmp#1,#2,{{\sl Rev.\ Mod.\ Phys.\/ }{\bf #1} (19#2)\ }
\def\sm#1,#2,{{\sl Selec. Math.\/ }{\bf #1} (19#2)\ }
\def\yadfiz#1,#2,#3[#4,#5]{{\sl Yad.\ Fiz.\/ }{\bf #1} (19#2) #3%
\ [{\sl Sov.\ J.\ Nucl.\ Phys.\/ }{\bf #4} (19#2) #5]}
\def\zh#1,#2,#3[#4,#5]{{\sl Zh.\ Exp.\ Theor.\ Fiz.\/ }{\bf #1} (19#2) #3%
\ [{\sl Sov.\ Phys.\ JETP\/ }{\bf #4} (19#2) #5]}

\begin {document}


\begin{titlepage}

\begin{center}
\today
\hfill LBNL-41481, UCB-PTH-98/14, RU-98-05 \\
\hfill                  hep-th/9803051

\vskip 1.5 cm
{\large \bf M Theory on $AdS_p\times S^{11-p}$
and Superconformal Field Theories}
\vskip 1 cm
{Ofer Aharony$^{1a}$,
Yaron Oz$^{2b}$,
Zheng Yin$^{2c}$}\\
\vskip 0.5cm
{\sl $^1$Department of Physics and Astronomy \\
Rutgers University, Piscataway, NJ 08855-0849, U.S.A. \\
and \\
Institute for Theoretical Physics	\\
University of California, Santa Barbara, CA 93106, U.S.A.}
\vskip 0.5cm
{\sl $^2$Department of Physics,
University of California at Berkeley\\
Berkeley, CA 94720-7300, U.S.A.\\
and\\
Theoretical Physics Group, Mail Stop 50A--5101\\
Ernest Orlando Lawrence Berkeley National Laboratory, \\
1 Cyclotron Road, Berkeley, CA 94720, U.S.A.\\
{\footnotesize email: $^a$ oferah@physics.rutgers.edu, $^b$ yaronoz@thsrv.lbl.gov, 
	$^c$ zyin@thsrv.lbl.gov.}}

\end{center}

\vskip 0.5 cm
\begin{abstract}

We study the large $N$ limit of the interacting superconformal field
theories associated with $N$ M5 branes or M2 branes using the recently
proposed relation between these theories and M theory on $AdS$
spaces. We first analyze the spectrum of chiral operators of the 6d
$(0,2)$ theory associated with M5 branes in flat space, and find full
agreement with earlier results obtained using its DLCQ description as
quantum mechanics on a moduli space of instantons.
We then perform a similar analysis for the $D_N$ type 6d $(0,2)$
theories associated with M5 branes at an $R^5/Z_2$ singularity, and
for the 3d ${\cal N}=8$ superconformal field theories associated with
M2 branes in flat space and at an $R^8/Z_2$ singularity
respectively. Little is known about these three theories, and our
study yields for the first time their spectrum of chiral operators (in
the large $N$ limit).

\end{abstract}
\end{titlepage}

\section{Introduction}

A duality between a certain limit of some superconformal field
theories (SCFTs) in $d$ dimensions and string or M theory compactified
on spaces of the form $AdS_{d+1} \times W$ has recently been proposed
in \cite{mal} (see also
\cite{DPS,kleb1,kleb2,kleb3,kleb4,MS,Polyakov}).  Here $W$ is a
compact manifold which in the maximally supersymmetric cases is a
sphere.  A precise correspondence between the supergravity limit on
the $AdS_{d+1}$ side and an appropriate large $N$ limit on the
conformal field theory side has been formulated in
\cite{GKP,witten-one}.  According to \cite{witten-one} the correlation
functions in the conformal field theory, which has as its spacetime
$M_d$, the boundary at infinity of $AdS_{d+1}$, can be calculated
systematically from the dependence of the supergravity action on the
asymptotic behaviour of its fields at the boundary $M_d$. In
particular, one can deduce the scaling dimensions of operators in the
conformal field theory from the masses of particles in string theory
(or M theory). Using this correspondence, the dimensions of chiral
fields in four dimensional ${\cal N}=4$ SYM were matched with the
masses of Kaluza-Klein states on $AdS_5 \times S^5$. Note that for
chiral primary fields (which are in short representations of the
superconformal algebra) the dimension is determined in terms of the
R-symmetry representation, and it cannot receive any corrections (see
\cite{seiberg,minwalla} for details). Related works which appeared
recently are \cite{0,1,2,3,4,KS,6,65,7,8,9,LNV,newKleb,AV,CCDFFT}.

In this paper we will study the proposed duality for several SCFTs.
They are all realized as the low energy theories on the branes of M
theory. The first theory is the six dimensional $(0,2)$
superconformal field theory on the worldvolume of $N$ parallel M5
branes.  This theory has a Matrix-like DLCQ description as a quantum
mechanics on the moduli space of instantons \cite{abkss,witten}.
Using this description chiral primary operators in the theory were
identified with compact cohomologies of the resolved moduli space of
instantons, and their spectrum was computed in \cite{abs}. We will
discuss also the $(0,2)$ $D_k$ theories which arise for 5-branes at
$R^5/Z_2$ singularities. These theories also have a DLCQ description
\cite{abkss}, but their spectrum of operators has not been computed
until now. The final theory we study is the three dimensional superconformal
field theory on the worldvolume of $N$ parallel and coincident M2
branes.  This is the strong coupling (infrared) limit of 3d ${\cal N}
= 8$ U(N) gauge theories.  Little is known about this theory and it
does not have a matrix description.

In \cite{mal} it was proposed that the SCFT of $N$ M5 branes is dual
to M theory on $AdS_7 \times S^4$, while the SCFT on $N$ M2
branes is dual to M theory on $AdS_4 \times S^7$.  Our aim is to study
these relations between eleven dimensional supergravity and M theory
on the $AdS$ spaces and  the large $N$ limit of the corresponding
superconformal
field theories.  In the
next section we will briefly review the precise correspondence between
the supergravity and SCFTs.  In section 3 we will consider the
$(0,2)$ theory on $N$ M5 branes, as well as the $D_N$ $(0,2)$ theory.
We will obtain the dimensions of the chiral operators of these theories
 from the masses of Kaluza-Klein modes of supergravity on
$AdS_7 \times S^4$, and find agreement
with the spectrum predicted using DLCQ. In section 4 we will consider
the three dimensional SCFT corresponding to the low energy theory of
$N$ M2 branes.  Using the masses of the Kaluza-Klein modes of
supergravity on $AdS_4 \times S^7$ we compute the spectrum
of chiral operators of this theory at large $N$. Section 5 is devoted
to a discussion.

\section{SCFT/$AdS$ correspondence}

In the following we will briefly review the SCFT/$AdS$ correspondence
proposed in \cite{GKP,witten-one}.  The boundary $M_d$ of $AdS_{d+1}$ is a
$d$-dimensional Minkowski space with points at infinity added.  The
symmetry group of $AdS_{d+1}$ is $SO(d,2)$. It is also the conformal
group on $M_d$.  The proposed duality relates
string theory (or M theory) on $AdS_{d+1}$ to the large $N$ limit of
some SCFTs on its boundary
$M_d$.  In the Euclidean version the boundary is $S^d$.
Consider for simplicity the maximally supersymmetric case, so that the
internal space is also a sphere.  Let $\phi$
be a scalar field on $AdS_{d+1}$ and $\phi_0$ its restriction to the
boundary $S^d$ (defined appropriately for massive fields in
\cite{witten-one}).  According to the SCFT/$AdS$ correspondence $\phi_0$
couples to a conformal field ${\cal O}$ on the boundary via $\int_{S^d}
\phi_0 {\cal O}$.  The proposed relation between the generating functional
$\langle exp \int_{S^d}\phi_0 {\cal O} \rangle_{SCFT}$ of
the SCFT on the boundary and the $AdS_{d+1}$ theory is \cite{witten-one}
\beq \langle exp \int_{S^d}\phi_0 {\cal O} \rangle_{SCFT} = Z_s(\phi_0) \comma
\eeq where $Z_s(\phi_0)$ is the supergravity (string/M theory) partition
function computed with boundary condition $\phi \sim \phi_0$ at infinity.

When $\phi$ has mass $m$ the corresponding operator
${\cal O}$ has conformal dimension $\Delta$ given by
\beq
m^2 = \Delta(\Delta-d) \stop
\label{dim}
\eeq
Irrelevant, marginal and relevant operators of the boundary theory
correspond to massive, massless and ``tachyonic'' modes in the supergravity
theory.  If a $p$-form $C$ on $AdS$ is coupled to a $d-p$ form
operator $\cal O$ on the boundary, then the relation between the mass of
$C$ and the conformal dimension of $\cal O$ is given by
\beq
m^2 = (\Delta+p)(\Delta+p-d) \stop
\label{dimp}
\eeq
The value of $m^2$ in this formula refers to the eigenvalue of the Laplace
operator on the $AdS$ space. In the supergravity literature, the values that
are usually quoted for $p$-forms are the eigenvalues ${\tilde m}^2$ of the
appropriate Maxwell-like operators. The relation of these to the dimension
is given by
\beq
{\tilde m}^2 = (\Delta-p)(\Delta+p-d) \stop
\label{dimpm}
\eeq

Some of these chiral fields are universal.
There is always a massless graviton in the $AdS$, which couples
to the stress energy tensor of the SCFT
(of dimension $\Delta=d$). If the internal space $W$ has continuous
rotational symmetry, there are also massless vector
fields in its adjoint representation, coupling to the R symmetry currents
of the SCFT (of dimension $\Delta=d-1$).

\section{$(0, 2)$ SCFTs in Six Dimensions}

Consider M theory on $AdS_7 \times S^4$ with a 4-form flux of $N$
quanta on $S^4$, and with the ``radii'' of the $AdS_7$ and $S^4$ being
$R_{AdS} = 2R_{S^4} = 2l_p(\pi N)^{1/3}$. Eleven dimensional
supergravity is applicable at energies much smaller than the Planck
scale $1/l_p$. For large $N$ this includes the energy range of the KK
modes, whose mass is of the order of $1/R_{AdS}$ (and we will measure
it in these units below). The bosonic symmetry of this
compactification of eleven dimensional supergravity is $SO(6,2) \times
SO(5)$.

In \cite{mal} it was proposed that the $(0,2)$ conformal theory, which
is the decoupled intrinsic theory on $N$ parallel M5
branes\footnote{See \cite{seiberg} for a discussion and references
concerning these SCFTs.}, is dual to M theory on the above background
in some appropriate sense.  The $SO(6,2)$ part of the symmetry of the
supergravity theory is the conformal group of the SCFT, which can be
thought of as living on the boundary of the $AdS$ space.  The $SO(5)$
part of the symmetry corresponds to the R symmetry of the
superconformal theory.

The Kaluza-Klein excitations of supergravity, in the maximally
supersymmetric cases, all fall
into small representations of supersymmetry (since
they contain no spins larger than 2). Thus, their mass formula is protected
 from quantum and string/M theory corrections.  According to the proposal
in \cite{witten-one}, they couple to
chiral fields of the SCFT on the boundary,
whose scaling dimensions are similarly protected from
quantum corrections. The spectrum of the Kaluza-Klein harmonics of
supergravity on $AdS_7 \times S^4$ was
analyzed in \cite{van}. There are three families of scalar
excitations. Two families contain states with only positive $m^2$ and
correspond only to irrelevant operators.  One family contains also states
with negative and zero $m^2$.  They fall into
the $k$-th order symmetric traceless
representation of $SO(5)$ with unit multiplicity.
Their masses are given by \footnote{The
overall coefficient of the mass formula depends on a parameter $e$ giving the
scale of the internal manifold $W$ and
used in the relation between the 4-form field strength and the
totally antisymmetric tensor.  Here it is determined by matching the mass
formula and (\ref{dim}).}
\beq
m^2 = 4k(k-3),~~~~~k=2,3, \ldots \stop
\eeq
The field corresponding to $k=1$ also appears in the supergravity, and
this is the singleton which may be gauged away except at the boundary
of the $AdS$ space and decouples from all other operators. In the
field theory we can identify it with the decoupled free center of mass
motion. This will be true in all the constructions of this type, and in
the rest of the paper we will only discuss the interacting fields.

Using (\ref{dim}), the dimensions of the corresponding
operators in the SCFT are
\beq
dim({\cal O}) = \{2k,~~k=2,3, \ldots\}
\label{dims}
\stop
\eeq
These are precisely the dimensions of the chiral primary operators
found in \cite{abs}, which parameterize the space of flat directions
$(R^5)^N/S_N$ in a ``gauge'' invariant way\footnote{
	Note that for finite $N$ the number of
	these fields is larger than the dimension
	of the moduli space, so these fields
	are not all independent on the moduli space.}.
Thus, this can be viewed as a test of the conjecture of \cite{mal}.
In the matrix description of the $(0,2)$ conformal theory these
operators correspond to compact cohomology elements of the resolved
moduli space of instantons, localized at the origin \cite{abs}. Note
that in \cite{abs} only those chiral fields whose scalars are in
totally symmetric traceless representations of $SO(5)$ were analyzed, but
the duality suggests that these are the only chiral fields that have
finite dimensions for large $N$. For large $N$ we find that the field
with $k=2$ is the only relevant scalar deformation of the SCFT. This
deformation breaks supersymmetry completely, and it would be
interesting to analyze which non-supersymmetric field theory it leads
to in the infrared.

There is one family of vector
bosons that contains also massless states
\beq
{\tilde m}^2 = 4(k^2-1),~~~~~k=1,2, \ldots
\stop
\label{vp}
\eeq
Using (\ref{dimpm}), the dimensions of the corresponding
1-form operators in the SCFT are
\beq
dim({\cal O}) = \{2k+3,~~k=1,2, \ldots\}
\label{dimsp}
\stop
\eeq
The massless vector at $k=1$ in (\ref{vp}) corresponds to the dimension
five R symmetry current.

In general, chiral fields corresponding to all the towers of
Kaluza-Klein harmonics are related to the scalar operators of
(\ref{dims}) by the superconformal algebra, as discussed in the four
dimensional case in \cite{FFZ}. Each value of $k$ gives rise (at least
for large enough $k$) to one field in each tower of KK states, with an
$SO(5)$ representation that is determined by the representation of the
scalar field. In particular, the R symmetry currents and the stress
energy tensor sit in the same superconformal representation as the
scalar field with $k=2$ mentioned above. The highest component (in the
$\theta$ expansion) of these superconformal multiplets gives a series
of scalar operators whose dimension starts at 12. The lowest one of
these operators couples to the trace of the graviton in spacetime, as
discussed in \cite{newKleb}\footnote {Similar operators with dimension
$\Delta=2d$ exist also for $d=3$ and $d=4$.}.  In a superfield
notation, the tower of scalars (\ref{dims}) corresponds to $\theta^0$
terms in the multiplets, $\theta^2$ terms lead to a vector and a
self-dual 3-form, $\theta^4$ terms lead to a graviton, a scalar and a
2-form, $\theta^6$ terms lead again to a vector and a self-dual
3-form, and the $\theta^8$ terms lead again to a scalar. By $\theta^i$
terms here we mean fields which can be reached by acting with $i$ SUSY
generators on the scalars of (\ref{dims}). The terms with an odd
number of $\theta$'s include spinors and gravitinos.  For low values
of $k$, some of the terms with a higher number of $\theta$'s are
descendants of the terms with a lower number of $\theta$'s \cite{FFZ},
but for large $k$ all the fields in the multiplet are independent.

A simple generalization of this construction gives the large $N$ limit
of the $D_N$ $(0,2)$ SCFTs, which correspond to the low energy
theories of $N$ M5-branes coinciding at an $R^5/Z_2$ orientifold
singularity \cite{Witten0,DM}. In the original theory of the 5-branes,
the $Z_2$ acts by a reflection of the 5 directions transverse to
the 5-branes, and also by changing the sign of the 3-form field $C$ of
eleven dimensional supergravity. Taking the near horizon limit as in
\cite{mal}, we find that the $Z_2$ acts by a total inversion of the
$5$ Cartesian coordinates embedding $S^4$ in $R^5$ around the center
of the sphere. It has no fixed points so that the resulting manifold
is completely smooth (orbifolds in string theory were discussed in a
similar context in \cite{KS,LNV}). In the supergravity solution, all
we need to do therefore is to identify the fields on one side of the
sphere with the fields at the antipodal points, with also a sign
change for the $C$-field. For large $N$, when the sphere is large and
the antipodal points are far away from each other, we expect to still
be able to trust the supergravity solution after this
identification. The identification projects out half of the spherical
harmonics on the $S^4$. For scalars, only those with even $k$ in
(\ref{dims}) remain.  As before, the rest of the chiral operator
spectrum is determined by the superconformal symmetry. The correlation
functions of the remaining operators will be the same as they were,
with corresponding operators, in the $A_{2N}$ case (to leading order
in $1/N$). Note that these theories also have a DLCQ description
\cite{abkss} as a quantum mechanics on the moduli space of
$D_N$ instantons, but it is difficult to use it to compute the
spectrum since there is no obvious resolution of the singularities in
the moduli space (unlike the case discussed in \cite{abs}). Note also
that the orientifold carries a 5-brane charge of $(-\frac{1}{2})$, but
this becomes negligible in the large $N$ limit where we can trust the
supergravity solution.

\section{Three Dimensional ${\cal N}=8$ SCFTs}

Consider now M theory on $AdS_4 \times S^7$ with a 7-form flux of $N$
quanta on $S^7$.  The ``radii'' of $AdS_4$ and $S^7$ are given by
$2R_{AdS} = R_{S^7} = l_p(32 \pi^2 N)^{1/6}$.  Eleven dimensional
supergravity is applicable for energies of the order of $1/R_{AdS}$ if
$N$ is large. The bosonic symmetries of this compactification are
$SO(3,2) \times SO(8)$.

In \cite{mal} it was proposed that the conformal theory on $N$
parallel M2 branes on the boundary on $AdS_4$ is dual to M theory on
the above background\footnote{See \cite{seiberg} for a discussion and
references related to this conformal field theory.}.  The $SO(3,2)$ part
of the symmetry of the supergravity theory is the conformal group of
the SCFT on the boundary.  The $SO(8)$ part of the symmetry
corresponds to the R symmetry of the boundary SCFT.  From the point of
view of type IIA string theory, this SCFT is the infrared (strong
coupling) limit of the 3d ${\cal N}=8$ $U(N)$ gauge theory on $N$
coincident D2-branes.

As before, we study the correspondence between the Kaluza-Klein
excitations of supergravity and the chiral fields of the SCFT.
The spectrum of the
Kaluza-Klein harmonics of eleven dimensional supergravity on $AdS_4
\times S^7$ was analyzed in \cite{BCERS,van1}.  There are three
families of scalar excitations and two families of pseudoscalar
excitations. Three of them contain states with only positive $m^2$ and
correspond to irrelevant operators.  One family
contains also states with negative and zero $m^2$ with masses given by
\footnote{Note that to match our formulas with the conventions of
\cite{BCERS,van1}, there is (besides the overall normalization of the
mass as before) also a shift in $m^2$, which is shifted relative to
the Laplacian in \cite{BCERS,van1}.}
\beq
m^2 = \frac{1}{4}\left((k-2)(k-4)-8 \right) = \frac{1}{4} k (k-6),
	~~~~~k=2,3,...
\eeq
They fall into the $k$-th order symmetric traceless
representation of $SO(8)$ with unit multiplicity.

Using (\ref{dim}), the scaling dimensions of
the corresponding chiral operators in the SCFT
are
\beq
dim({\cal O}) = \{\frac{k}{2},~~k=2,3,...\}
\label{d}
\stop
\eeq
As before, we can identify these operators with the natural
gauge invariant coordinates on the moduli space of these theories, which is
$(R^8)^N/S_N$. Regarding this theory as the IR limit of the 3D
${\cal N}=8$ SYM theory, some of these operators may be identified with
operators of the form $\tr(X^{i_1} X^{i_2} ... X^{i_k})$, where the
$X^i$ are the scalar fields in the vector multiplet. However, in the
gauge theory there are only 7 scalar fields, and the additional field
arises from dualizing the vector field, which one cannot do explicitly
in the non-Abelian case. For $k=2 \dots 5$ these are relevant operators
in the conformal theory, and for $k=6$ they are marginal.

As before, chiral operators corresponding to the other towers of
Kaluza-Klein harmonics are related to those  from this family by the
superconformal symmetry. Again, we can identify the operators of
(\ref{d}) with $\theta^0$ components of the small superconformal
multiplet, and then the $\theta^2$ terms include a pseudoscalar and a
vector field, the $\theta^4$ terms include a graviton, a scalar and an
axial vector, the $\theta^6$ terms include a pseudoscalar and a vector
field, and the $\theta^8$ terms give a scalar. The odd $\theta$ terms
give spinors and gravitinos.

In this case, unlike the previous case, there is one other family of
pseudoscalar excitations which also contains states with
negative and zero $m^2$, corresponding to relevant and marginal operators
(respectively) in the SCFT.
The masses of this family are given by
\beq
m^2 = \frac{1}{4}\left ((k-1)(k+1) -8 \right ),~~~~~k=1,2,...  \stop
\label{mm}
\eeq
The $k$'th state transforms in a representation of $SO(8)$
corresponding to the product of a ${\bf 35}_c$ with $(k-1)$ ${\bf
8}_v$'s (in a symmetric traceless way).  
Using (\ref{dim}) the dimensions of the corresponding
operators in the SCFT are
\beq
dim({\cal O}) = \{\frac{k+3}{2},~~k=1,2,...\}
\label{dp}
\stop
\eeq
For instance, for $k=1$ we have 35 pseudoscalar relevant operators of
dimension 2.  In the UV SYM which flows to this SCFT, we can identify
these operators with a product of two fermions times $k-1$ scalars, as
in \cite{witten-one,FFZ}. The deformation of the SCFT by any of the
relevant scalar operators breaks the supersymmetry completely.
 
As in the M5 branes case, we identify one family of vector
bosons that also contains massless states. The masses of this family 
are given by
\beq
{\tilde m}^2 = \frac{1}{4}(k^2-1),~~~~~k=1,2,...
\stop
\label{vp2}
\eeq
Using (\ref{dimpm}), the dimensions of the corresponding
1-form operators in the SCFT are
\beq
dim({\cal O}) = \{\frac{k+3}{2},~~k=1,2,...\}
\label{dimsp2}
\stop
\eeq
The massless vector at $k=1$ in (\ref{vp2}) corresponds to the dimension
two R symmetry current.

As earlier, we can also put the M2 branes at an $R^8/Z_2$ singularity
(which this time is just an orbifold point).  Again we find that the
resulting 3D ${\cal N}=8$ SCFT is a truncation of the theory discussed
above.  Only half of its chiral operators remain, the even $k$ elements
of (\ref{d}) and the operators related to them by superconformal symmetry.

\section{Discussion}

In this note we computed the spectrum of chiral operators in the large
$N$ limit of several series of supersymmetric conformal
field theories with 16 supercharges using the conjecture of \cite{mal}.
In general, on the supergravity
side the only operators of low $m^2$ are the KK modes, which fall into
small representations.  Thus, the conjecture of \cite{mal} implies
that only chiral operators of the SCFT have
dimensions that do not grow with $N$ in the large $N$ limit. This is
quite surprising from the conformal field theory point of view, as is
the similar statement for the large $g^2N$ limit of the 4D ${\cal N}=4$ SYM
theories. In the supergravity approximation we cannot reliably study
non-chiral operators at all. In principle, in the full M theory such
operators could be analyzed. We would expect the $m^2$ of such states to be
of the order of $M_p^2$, which translates into operators of dimension
$N^{1/6}$ in the six dimensional case and $N^{1/3}$ in the three
dimensional case. However, it is difficult to access such operators
 from the M theory side, since there is
not yet any known non-perturbative definition for M theory on constant
curvature spaces (it is not clear how to generalize the DLCQ
formulation of Matrix theory \cite{BFSS} to this case).

In addition to the conjecture about the large $N$ behavior of the
superconformal field theories in \cite{mal}, there was also a
conjecture about the agreement of the finite $N$ theories with the
appropriate string/M theory backgrounds. However, it is not clear how
to check this (very strong) conjecture. One obvious property of the
finite $N$ theories is that their spectrum of (``single-particle'')
chiral primary operators is finite, namely the series of operators
described above are truncated at $k=N$. This corresponds to a
dimension of order $N$, which in turn corresponds to an $m^2$ that is
much larger than the Planck scale, so it is not clear how to see this
in the string/M theory side. Presumably, in order to check this
conjecture in some way, one would need to compute higher order
corrections in $1/N$ (say, to 3-point functions). In the case of the
6D theories, such computation are in principle accessible on the SCFT
side through their DLCQ construction \cite{abs}, but it is not clear
how to compute such quantum corrections on the M theory side.

\section*{Acknowledgements}

We would like to thank K. Bardakci and S. Kachru for discussions. O.A.
would like to thank J. M. Maldacena and the other participants of the
``Duality in String Theory'' workshop at the ITP for useful
discussions.  The work of Y.O. and Z.Y. was supported in part by NSF
grant PHY-951497 and DOE grant DE-AC03-76SF00098.  The work of
O.A. was supported in part by the National Science Foundation under
grant no. PHY94-07194, and by the DOE under grant
no. \#DE-FG02-96ER40559.

\end{document}